\documentclass[tightenlines,superscriptaddress,eqsecnum,floats,nofootinbib,showpacs]
{revtex4}
\usepackage{amssymb}
\usepackage{stmaryrd}
\usepackage{amsmath}
\usepackage{amsfonts}
\usepackage{mathrsfs}

\usepackage{amsmath,amssymb,amsfonts}
\usepackage{graphicx}
\def\be{\begin{equation}}
\def\ee{\end{equation}}
\def\ba{\begin{eqnarray}}
\def\ea{\end{eqnarray}}
\def\nn{\nonumber}

\newcommand{\abs}[1]{{\left|{#1}\right|}} % abs (variant delimiters)
\newcommand{\kt}{{\tilde{K}}} % K tilde
\newcommand{\R}{\mathcal {R}} % 4-d curvature scalar
\newcommand{\ints}{{\int_\Sigma}} % integral

\newcommand{\Tr}{\mathrm{Tr}} % trace
\newcommand{\grav}{\mathrm{gr}} % gravitational part
\newcommand{\sca}{\mathrm{sc}} % scalar part
\newcommand{\kin}{\mathrm{kin}} % kinematic (Hilbert space, states)
\newcommand{\hil}{\mathcal{H}} % Hilbert space
\newcommand{\Euc}{H^{E}} % Euclidean scalar

\begin{document}

%\date\today
%\preprint{????}

\title{Nonperturbative Loop Quantization of Scalar-Tensor Theories of Gravity}

\author{Xiangdong Zhang}\email{zhangxiangdong@mail.bnu.edu.cn}
\author{Yongge Ma\footnote{Corresponding author}}\email{mayg@bnu.edu.cn}

\affiliation{Department of Physics, Beijing Normal University,
Beijing 100875, China}

\begin{abstract}
The Hamiltonian formulation of scalar-tensor theories of gravity is
derived from their Lagrangian formulation by Hamiltonian analysis.
The Hamiltonian formalism marks off two sectors of the theories by
the coupling parameter $\omega(\phi)$. In the sector of
$\omega(\phi)=-\frac{3}{2}$, the feasible theories are restricted
and a new primary constraint generating conformal transformations of
spacetime is obtained, while in the other sector of
$\omega(\phi)\neq-\frac{3}{2}$, the canonical structure and
constraint algebra of the theories are similar to those of general
relativity coupled with a scalar field. By canonical
transformations, we further obtain the connection dynamical
formalism of the scalar-tensor theories with real
$su(2)$-connections as configuration variables in both sectors. This
formalism enables us to extend the scheme of non-perturbative loop
quantum gravity to the scalar-tensor theories. The quantum
kinematical framework for the scalar-tensor theories is rigorously
constructed. Both the Hamiltonian constraint operator and master
constraint operator are well defined and proposed to represent
quantum dynamics. Thus loop quantum gravity method is also valid for
general scalar-tensor theories.

\end{abstract}


\maketitle

\section{Introduction}

In the past 25 years, loop quantum gravity(LQG), a background
independent approach to quantize general relativity (GR), has been
widely investigated \cite{Ro04,Th07,As04,Ma07}. It is remarkable
that, as a non-renormalizable theory, GR can be non-perturbatively
quantized by the loop quantization procedure. This
background-independent quantization method relies on the key
observation that classical GR can be cast into the connection
dynamical formalism with structure group of $SU(2)$. Thus one is
naturally led to ask whether GR is a unique relativistic theory of
gravity with connection dynamical character. It was recently shown
in Refs.\cite{Zh11,Zh11b} that metric $f(\R)$ theories can also be
cast into the connection dynamical formalism, and hence LQG has been
extended to metric $f(\R)$ theories. In fact, modified gravity
theories have recently received increased attention in issues
related to "dark Universe" and non-trivial tests on gravity beyond
GR. Besides $f(\R)$ theories, a well-known competing relativistic
theory of gravity was proposed by Brans and Dicke in 1961 \cite{BD},
which is apparently compatible with Mach's principle. To represent a
varying "gravitational constant", a scalar field is non-minimally
coupled to the metric in Brans-Dicke theory. To interpret the
observational results within the framework of a broad class of
theories, the Brans-Dicke theory was generalized by Bergmann
\cite{bergmann} and Wagoner \cite{wagoner} to general scalar-tensor
theories (STT). Scalar-tensor modifications of GR have also become
very popular in unification schemes such as string theory (see e.g.
\cite{tayler,maeda,damour}). On the other hand, since 1998, a series
of independent observations, including type Ia supernova, weak lens,
cosmic microwave background anisotropy, baryon oscillation, etc,
implied that our universe is currently undergoing a period of
accelerated expansion\cite{Fr08}. These results have caused the
"dark energy" problem in the framework of GR. It is reasonable to
consider the possibility that GR is not a valid theory of gravity on
a galactic or cosmological scale. The scalar field in STT of gravity
is then expected to account for "dark energy", since it can
naturally lead to cosmological acceleration in certain models (see
e.g. \cite{Boiss00,Banerjee,Qiang,Boiss11}). Moreover, some models
of STT of gravity may account for the "dark matter" problem
\cite{lee,catena,kim}, which was revealed by the observed rotation
curve of galaxy clusters.

A large part of the non-trivial tests on gravity theory is related
to Einstein's equivalence principle (EEP) \cite{will}. There exist
many local experiments in solar-system supporting EEP, which implies
the metric theories of gravity. Actually, STT are a class of
representative metric theories, which have been received most
attention. Note that the metric $f(\R)$ theories and Palatini
$f(\R)$ theories are equivalent to the special kind of STT with the
coupling parameter $\omega(\phi)=0$ and $\omega(\phi)=-\frac32$
respectively\cite{So}, while the original Brans-Dicke theory is the
particular case of constant $\omega$ and vanishing potential of
$\phi$. Thus it is interesting to see whether this class of metric
theories of gravity could be quantized nonperturbatively. In this
paper, we will first do Hamiltonian analysis of the STT of gravity.
Based on the resulted connection dynamical formalism, we then
quantize the STT by extending the nonperturbative quantization
procedure of LQG in the way similar to loop quantum $f(\R)$ gravity
\cite{Zh11,Zh11b}. Nevertheless, the STT that we are considering are
a much more general class of metric theories of gravity than metric
$f(\R)$ theories . Throughout the paper, we use Greek alphabet for
spacetime indices, Latin alphabet a,b,c,..., for spatial indices,
and i,j,k,..., for internal indices.

\section{Hamiltonian analysis}

The most general action of STT reads
\ba S(g)=\int_\Sigma
d^4x\sqrt{-g}[\frac12(\phi\R-\frac{\omega(\phi)}{\phi}(\partial_\mu\phi)\partial^\mu\phi)-\xi(\phi)]\label{action}
\ea
where we set $8\pi G=1$, $\R$ denotes the scalar curvature of
spacetime metric $g_{\mu\nu}$, the coupling parameter $\omega(\phi)$
and potential $\xi(\phi)$ can be arbitrary functions of scalar field
$\phi$. Variations of the action (\ref{action}) with respect to
$g_{ab}$ and $\phi$ give equations of motion.
\ba &&\phi G_{\mu\nu}=\nabla_\mu\nabla_\nu\phi-g_{\mu\nu}\Box\phi
+\frac{\omega(\phi)}{\phi}[(\partial_\mu\phi)\partial_\nu\phi-\frac12g_{\mu\nu}(\nabla\phi)^2]-g_{\mu\nu}\xi(\phi),\label{01}
\\
&&\R+\frac{2\omega(\phi)}{\phi}\Box\phi-\frac{\omega(\phi)}{\phi^2}(\partial_\mu\phi)\partial^\mu\phi
+\frac{\omega'(\phi)}{\phi}(\partial_\mu\phi)\partial^\mu\phi-2\xi'(\phi)=0,\label{02}\ea
where a prime over a letter denotes a derivative with respect to the
argument,  $\nabla_\mu$ is the covariant derivative compatible with $g_{\mu\nu}$
and $\Box\equiv g^{\mu\nu}\nabla_\mu\nabla_\nu$. By doing 3+1 decomposition of the
spacetime, the four-dimensional (4d) scalar curvature can be
expressed as
\ba \mathcal
{R}=K_{ab}K^{ab}-K^2+R+\frac{2}{\sqrt{-g}}\partial_\mu(\sqrt{-g}n^\mu
K)-\frac{2}{N\sqrt{h}}\partial_a
(\sqrt{h}h^{ab}\partial_bN)\label{03} \ea
where $K_{ab}$ is the
extrinsic curvature of a spatial hypersurface $\Sigma$, $K\equiv
K_{ab}h^{ab}$, $R$ denotes the scalar curvature of the 3-metric
$h_{ab}$ induced on $\Sigma$, $n^\mu$ is the unit normal of $\Sigma$
and $N$ is the lapse function. By Legendre transformation, the
momenta conjugate to the dynamical variables $h_{ab}$ and $\phi$ are
defined respectively as
\ba p^{ab}&=&\frac{\partial\mathcal
{L}}{\partial\dot{h}_{ab}}=\frac{\sqrt{h}}{2}[\phi(K^{ab}-Kh^{ab})-\frac{h^{ab}}{N}(\dot{\phi}-N^c\partial_c\phi)], \label{04}\\
\pi&=&\frac{\partial\mathcal
{L}}{\partial\dot{\phi}}=-\sqrt{h}(K-\frac{\omega(\phi)}{N\phi}(\dot{\phi}-N^c\partial_c\phi)),\label{pi}
\ea
where $N^c$ is the shift vector. Combination the trace of Eq.
(\ref{04}) and Eq. (\ref{pi}) gives
\ba
(3+2\omega(\phi))(\dot{\phi}-N^a\partial_a\phi)=\frac{2N}{\sqrt{h}}(\phi\pi-p).\label{Sconstraint}
\ea
It is easy to see from Eq. (\ref{Sconstraint}) that one extra
constraint $S=p-\phi\pi=0$ emerges when $\omega(\phi)=-\frac32$.
Hence it is natural to mark off two sectors of the theories by
$\omega(\phi)\neq-\frac32$ and $\omega(\phi)=-\frac32$.

\subsection{Sector of $\omega(\phi)\neq -3/2$ }

In the case of $\omega(\phi)\neq -3/2$, the Hamiltonian of STT can
be derived as a liner combination of constraints as
\ba H_{total}=\int_\Sigma d^3x(N^aV_a+NH),\label{htotal} \ea
where
the smeared diffeomorphism and Hamiltonian constraints read
respectively
\ba V(\overrightarrow{N})&=&\int_\Sigma d^3xN^aV_a =\int_\Sigma
d^3xN^a\left(-2D^b(p_{ab})+\pi\partial_a\phi\right),\label{dc}\\
H(N)&=&\int_\Sigma d^3xNH \nn\\
&=&\int_\Sigma
d^3xN\left[\frac2{\sqrt{h}}\left(\frac{p_{ab}p^{ab}-\frac12p^2}{\phi}+\frac{(p-\phi\pi)^2}{2\phi(3+2\omega)}\right)
+\frac12\sqrt{h}(-\phi R+\frac{\omega(\phi)}{\phi}(D_a\phi)
D^a\phi+2D_aD^a\phi+2\xi(\phi))\right].\label{hc}\nn\\
\ea
By the symplectic structure
\ba
\{h_{ab}(x),p^{cd}(y)\}&=&\delta^{(c}_a\delta^{d)}_b\delta^3(x,y),\nn\\
\{\phi(x),\pi(y)\}&=&\delta^3(x,y), \label{poission}\ea
lengthy but
straightforward calculations show that the constraints (\ref{dc})
and (\ref{hc}) comprise a first-class system similar to GR as:
\ba \{V(\overrightarrow{N}),V(\overrightarrow{N}^\prime)\}&=&V([\overrightarrow{N},\overrightarrow{N}^\prime]), \nn\\
\{H(M),V(\overrightarrow{N})\}&=&-H(\mathcal
{L}_{\overrightarrow{N}}M), \nn\\
\{H(N),H(M)\}&=&V(ND^aM-MD^aN). \ea
Note that the Hamiltonian
formulation of Brans-Dicke theory was studied in Ref.\cite{olmo}
(but the term $D_aD^a\phi$ of Hamiltonian constraint was missing
there). We now show that the above Hamiltonian formalism of STT is
equivalent to their initial value formalism obtained in
Ref.\cite{Sa}. In the Hamiltonian formalism, the evolution equations
of the basic canonical variables can be derived by taking Poisson
brackets with the Hamiltonian (\ref{htotal}). Thus it is
easy to check that the evolution equation of $h_{ab}$ is just the
definition of $K_{ab}$. Also we have
\ba
\dot{\phi}=\{\phi,H_{total}\}=\frac{2N}{(3+2\omega)\sqrt{h}}(\phi\pi-p)+N^a\partial_a\phi
\ea
which is nothing but  Eq.(\ref{Sconstraint}). Moreover, it is
easy to obtain
\ba \dot{\pi}
&=&\partial_a(N^a\pi)+\frac{N\sqrt{h}}{2}(K_{ab}K^{ab}-K^2)
+\frac{N\sqrt{h}}{2}R-\partial_a(\sqrt{h}h^{ab}\partial_bN)\nn\\
&+&\frac{N\omega\sqrt{h}}{2\phi^2}(D_a\phi) D^a\phi+\sqrt{h}\omega
D_a(\frac{N}{\phi}D^a\phi)-\frac{\omega\sqrt{h}}{2\phi^2N}(\dot{\phi}-N^c\partial_c\phi)^2
+\frac{\omega'(\phi)N\sqrt{h}}{2\phi}(D_a\phi)
D^a\phi-N\sqrt{h}\xi'(\phi). \ea
Using Eqs. (\ref{03}), (\ref{pi})
and $n^0=\frac1N,n^a=-\frac{N^a}{N}, \sqrt{-g}=N\sqrt{h}$, we can
get
\ba \dot{\pi} -\partial_a(N^a\pi)-\partial_\mu(\sqrt{-g}n^\mu
K)&=&
\partial_\nu(\frac{\sqrt{-g}\omega(\phi)}{\phi}n^\nu
n^\sigma\partial_\sigma\phi)\nn\\
&=&\frac{\sqrt{-g}\R}{2}
-\frac{\sqrt{-g}\omega}{2\phi^2}(n^\sigma\partial_\sigma\phi)^2
-\frac{\sqrt{-g}\omega}{2\phi^2}(D_a\phi)
D^a\phi\nn\\
&+&\frac{\omega}{\phi}\partial_a(\sqrt{-g}h^{ab}\partial_b\phi)+\frac{\omega'(\phi)N\sqrt{h}}{2\phi}(D_a\phi)
D^a\phi-\sqrt{-g}\xi'(\phi), \ea
which is equivalent to Eq. (\ref{02}). On the other hand, in the
Lagrangian formalism, the 00-component of Eq.(\ref{01}) reads
\ba
\phi G_{\mu\nu}n^\mu
n^\nu&=&\frac\phi2(R-K_{ab}K^{ab}+K^2)\nn\\
&=&D_aD^a\phi-\frac
KN(\dot{\phi}-N^c\partial_c\phi)+\frac{\omega}{2\phi}(D_a\phi)
D^a\phi +\frac{\omega}{2\phi}(n^\mu\partial_\mu\phi)^2+\xi(\phi),
\ea
where the fact $g_{\mu\nu}n^\mu n^\nu=-1$, $h^{\mu\nu}n_\nu=0$
and
$n^\sigma\partial_\sigma\phi=\frac1N(\dot{\phi}-N^c\partial_c\phi)$
have been used in the above derivation. Note that the Hamiltonian
constraint in (\ref{hc}) can be expressed as
\ba 0=H
&=&\frac{\sqrt{h}\phi}{2}(K_{ab}K^{ab}-K^2-R)+\frac{\sqrt{h}}{2}(\frac{\omega}{\phi}(D_a\phi)
D^a\phi+2D_aD^a\phi)\nn\\
&-&\sqrt{h}\frac KN(\dot{\phi} -N^c\partial_c\phi)
+\frac{\omega\sqrt{h}}{2\phi}(n^\mu\partial_\mu\phi)^2+\sqrt{h}\xi(\phi).
\ea
So the 00-component of (\ref{02}) is equivalent to the
Hamiltonian constraint. Since
\ba \phi G_{\mu\nu}n^\mu
h^\nu_b=\phi(D_aK^a_b-D_bK)  \quad and \quad g_{\mu\nu}n^\mu
h^\nu_b=0,\ea
we can get from the 0a-components of (\ref{01}) that
\ba 0=D_a(\phi K^a_b)-D_b(\phi K)+KD_b\phi
-D_b(\frac{1}{N}(\dot{\phi}-N^c\partial_c\phi))-(n^\mu\partial_\mu\phi)\frac{\omega}{\phi}D_b\phi
\nn\\
=\frac{2}{\sqrt{h}}
D_a\left(\frac{\sqrt{h}}{2}[\phi(K^a_b-Kh^a_b)-\frac{h^a_b}{N}(\dot{\phi}-N^c\partial_c\phi)]\right)
-\frac{\pi}{\sqrt{h}}\partial_b\phi. \ea
This is nothing but the
diffeomorphism constraints in (\ref{dc}). Now we turn to the
ab-components of (\ref{01}), we will show that they are equivalent
to the equation of motion of $p_{ab}$, which reads
\ba
\dot{p}_{ab}&=&\frac{h_{ab}N}{\sqrt{h}}(\frac{p_{cd}p^{cd}-\frac12p^2}{\phi}+\frac{(p-\phi\pi)^2}{2\phi(3+2\omega)})
+\frac{4N}{\sqrt{h}}(\frac{p_{ac}p^c_b-\frac12pp_{ab}}{\phi}+\frac{(p-\phi\pi)p_{ab}}{2\phi(3+2\omega)})\nn\\
&+&\frac N4\sqrt{h}h_{ab}\phi R-\frac N2\sqrt{h}\phi R_{ab}-\frac
N2\sqrt{h}h_{ab}D_cD^c\phi-D_{(a}N\sqrt{h}D_{b)}\phi-\frac
{N\omega}{4\phi}\sqrt{h}h_{ab}D_c\phi D^c\phi+\frac
{N\omega}{2\phi}\sqrt{h}D_a\phi D_b\phi \nn\\
&+&\frac{\sqrt{h}}{2}(D_{(a}D_{b)}(N\phi)-h_{ab}D_cD^c(N\phi))+2p_{c(a}D^cN_{b)}+D_c(p_{ab}N^c)
-\frac12N\sqrt{h}h_{ab}\xi(\phi).\label{pab} \ea
Since the initial
value formalism of STT has been obtained in Ref.\cite{Sa}, we may
use Eq.(\ref{pab}) to derive the time derivative of the extrinsic
curvature:
\ba K_{ab}
&=&\frac{2(p_{ab}-\frac{1+\omega}{3+2\omega}ph_{ab})}{\phi\sqrt{h}}-\frac{\pi
h_{ab}}{(3+2\omega)\sqrt{h}}. \ea
Straightforward calculations give
\ba \dot{K}_{ab}&=&2NK_{ac}K^c_b-NKK_{ab}+\mathcal
{L}_{\overrightarrow{N}}K_{ab}-NR_{ab}+D_aD_bN+\frac{N}{\phi}D_aD_b\phi\nn\\
&+&\frac{N\omega}{\phi^2}(D_a\phi)
D_b\phi-\frac{n^\sigma\partial_\sigma\phi}{\phi}NK_{ab}+Nh_{ab}(\frac12\Box\phi+\frac{\xi(\phi)}{\phi}).\label{kab}
\ea
It is easy to see that Eq.(\ref{kab}) is equivalent to Eq. (69)
in Ref.[24]. Note that there is a sign difference between the
definition of our extrinsic curvature and that in Ref.\cite{Sa}. To
summarize, we have shown that the Hamiltonian formalism of STT with $\omega(\phi)\neq -3/2$ is
equivalent to their Lagrangian formalism.

Since the geometric canonical variables of STT in this sector are as
same as those of metric $f(\R)$ theories \cite{Zh11b}, we can use
the same canonical transformations of $f(\R)$ theories to obtain the
connection dynamical formalism of the STT. Let
\ba\tilde{K}^{ab}=\phi
K^{ab}+\frac{h^{ab}}{2N}(\dot{\phi}-N^c\partial_c\phi)=\phi
K^{ab}+\frac{h^{ab}}{(3+2\omega)\sqrt{h}}(\phi\pi-p).\ea The new
canonical geometric variables are \ba E^a_i=\sqrt{h}e^a_i,\quad
A^i_a=\Gamma^i_a+\gamma\tilde{K}^i_a,\label{newvaribles}\ea where
$e^a_i$ is the triad such that $h_{ab}e^a_ie^b_j=\delta_{ij}$,
$\tilde{K}^a_i\equiv\tilde{K}^{ab}e_b^i$, $\Gamma^i_a$ is the spin
connection determined by $E^a_i$, and $\gamma$ is a nonzero real
number. It is clear that our new variable $A^i_a$ coincides with the
Ashtekar-Barbero connection \cite{As86,Ba} when $\phi=1$. The
Poisson brackets among the new variables read: \ba
\{A^j_a(x),E_k^b(y)\}&=&\gamma\delta^b_a\delta^j_k\delta(x,y),\nn\\
\{A_a^i(x),A_b^j(y)\}&=&0,\quad \{E_j^a(x),E_k^b(y)\}=0. \ea
Now,
the phase space of the STT consists of conjugate pairs $(A_a^i,E^b_j)$
and $(\phi,\pi)$, with the additional Gaussian constraint
\ba \mathcal
{G}_i=\mathscr{D}_aE^a_i\equiv\partial_aE^a_i+\epsilon_{ijk}A^j_aE^{ak},
\label{GC}\ea
which justifies $A^i_a$ as an $su(2)$-connection. Note
that, had we let $\gamma=\pm i$, the (anti-)self-dual complex
connection formalism would be obtained. The original vector and
Hamiltonian constraints can be respectively written up to Gaussian
constraint as
\ba V_a &=&\frac1\gamma F^i_{ab}E^b_i+\pi\partial_a\phi,
\label{diff}\ea

\ba H
&=&\frac{\phi}{2}\left[F^j_{ab}-(\gamma^2+\frac{1}{\phi^2})\varepsilon_{jmn}\tilde{K}^m_a\tilde{K}^n_b\right]
\frac{\varepsilon_{jkl}
E^a_kE^b_l}{\sqrt{h}}\nn\\
&+&\frac1{3+2\omega(\phi)}\left(\frac{(\tilde{K}^i_aE^a_i)^2}{\phi\sqrt{h}}+
2\frac{(\tilde{K}^i_aE^a_i)\pi}{\sqrt{h}}+\frac{\pi^2\phi}{\sqrt{h}}\right) \nn\\
&+&\frac{\omega(\phi)}{2\phi}\sqrt{h}(D_a\phi) D^a\phi+\sqrt{h}D_aD^a\phi+\sqrt{h}\xi(\phi),\label{hamilton}
\ea
where
$F^i_{ab}\equiv2\partial_{[a}A^i_{b]}+\epsilon^i_{kl}A_a^kA_b^l$ is
the curvature of $A_a^i$. The total Hamiltonian can be expressed as
a linear combination
\ba H_{total}=\ints\Lambda^i\mathcal
{G}_i+N^aV_a+NH.\ea
It is easy to check that the smeared Gaussian
constraint, $\mathcal {G}(\Lambda):=\int_\Sigma
d^3x\Lambda^i(x)\mathcal {G}_i(x)$, generates $SU(2)$ gauge
transformations on the phase space, while the smeared constraint
$\mathcal {V}(\overrightarrow{N}):=\int_\Sigma
d^3xN^a(V_a-A_a^i\mathcal {G}_i)$ generates spatial
diffeomorphism transformations on the phase space. Together with the
smeared Hamiltonian constraint $H(N)=\int_\Sigma d^3xNH$, we can
show that the constraints algebra has the following form:
\ba
\{\mathcal {G}(\Lambda),\mathcal {G}(\Lambda^\prime)\}&=&\mathcal
{G}([\Lambda,\Lambda^\prime]),\label{eqsA} \\
\{\mathcal
{G}(\Lambda),\mathcal{V}(\overrightarrow{N})\}&=&-\mathcal{G}(\mathcal
{L}_{\overrightarrow{N}}\Lambda,),\\
\{\mathcal {G}(\Lambda),H(N)\}&=&0,\\
\{\mathcal {V}(\overrightarrow{N}),\mathcal
{V}(\overrightarrow{N}^\prime)\}&=&\mathcal
{V}([\overrightarrow{N},\overrightarrow{N}^\prime]), \\
\{\mathcal {V}(\overrightarrow{N}),H(M)\}&=&H(\mathcal
{L}_{\overrightarrow{N}}M),\label{eqsE}\\
\{H(N),H(M)\}&=&\mathcal {V}(ND^aM-MD^aN)\nn\\
&+&\mathcal
{G}\left((N\partial_aM-M\partial_aN)h^{ab}A_b\right)\nn\\
&-&\frac{[E^aD_aN,E^bD_bM]^i}{h}\mathcal {G}_i\nn\\
&-&\gamma^2\frac{[E^aD_a(\phi N),E^bD_b(\phi M)]^i}{h}\mathcal
{G}_i.\label{eqsb}\ea
One may understand Eqs.(\ref{eqsA}-\ref{eqsE}) by the geometrical interpretations of $\mathcal {G}(\Lambda)$ and
$\mathcal {V}(\overrightarrow{N})$. The detail calculation on the
Poisson bracket (\ref{eqsb}) between the two smeared Hamiltonian constraints will
be presented in the Appendix. Hence the constraints are of first class.
Moreover, the constraint algebra of GR can be recovered for the
special case when $\phi=1$. To summarize, the STT of gravity in the
sector $\omega(\phi)\neq -3/2$ have been cast into the
$su(2)$-connection dynamical formalism. The resulted Hamiltonian
structure is similar to metric $f(R)$ theories.

\subsection{Sector of $\omega(\phi)= -3/2$ }

In the special case of $\omega(\phi)= -3/2$, Eq. (\ref{Sconstraint})
implied an extra primary constraint $S=0$, which we call "conformal"
constraint. Hence, as pointed out in Ref.\cite{olmo}, the total
Hamiltonian in this case can be expressed as a liner combination of
constraints as
\ba H_{total}=\int_\Sigma d^3x(N^aV_a+NH+\lambda S),\label{htotal1}
\ea
where the smeared diffeomorphism $V(\overrightarrow{N})$ is as
same as (\ref{dc}), while the Hamiltonian and conformal constraints
read respectively
\ba
H(N)&=&\int_\Sigma d^3xNH \nn\\
&=&\int_\Sigma
d^3xN\left[\frac2{\sqrt{h}}\left(\frac{p_{ab}p^{ab}-\frac12p^2}{\phi}\right)
+\frac12\sqrt{h}(-\phi R-\frac{3}{2\phi}(D_a\phi)
D^a\phi+2D_aD^a\phi+2\xi(\phi))\right],\label{hc1}\\
S(\lambda)&=&\int_\Sigma d^3x\lambda S=\int_\Sigma
d^3x\lambda(p-\phi\pi).\label{sc}\ea
By the symplectic structure
(\ref{poission}), detailed calculations show that
\ba \{H(M),V(\overrightarrow{N})\}&=&-H(\mathcal
{L}_{\overrightarrow{N}}M),\quad \{S(\lambda),V(\overrightarrow{N})\}=-S(\mathcal
{L}_{\overrightarrow{N}}\lambda),\label{VHS}\\
\{H(N),H(M)\}&=&V(ND^aM-MD^aN)+S(\frac{D_a\phi}{\phi}(ND^aM-MD^aN)),\label{HH}\\
\{S(\lambda),H(M)\}&=&H(\frac{\lambda M}{2})+\ints
N\lambda\sqrt{h}(-2\xi(\phi)+\phi\xi'(\phi)).\label{Sc} \ea
One may understand Eqs.(\ref{VHS}) by the geometrical interpretations of
$\mathcal {V}(\overrightarrow{N})$. The detail calculations on the
Poisson brackets (\ref{HH}) and (\ref{Sc}) will
be presented in the Appendix.
The
Poisson bracket (\ref{Sc}) implies that, in order to maintain the
constraints $S$ and $H$ in the time evolution, we have to impose a
secondary constraint
\ba -2\xi(\phi)+\phi\xi'(\phi)=0.
\label{equationofV}\ea
It is easy to see that this constraint is of
second-class, and hence one has to solve it. As for the vacuum case
that we considered the solutions of Eq. (\ref{equationofV}) are
either $\xi(\phi)=0$ or $\xi(\phi)=C\phi^2$, where $C$ is certain
dimensional constant. Thus the consistency strongly restricted the
feasible STT in the sector $\omega(\phi)=-3/2$ to only two theories.
As pointed out in Ref.\cite{So}, for these two theories, the action
(\ref{action}) is invariant under the following conformal
transformations:
\ba g_{\mu\nu}\rightarrow e^\lambda
g_{\mu\nu},\quad \phi\rightarrow e^{-\lambda}\phi.\label{conformalt}
\ea
Hence, besides diffeomorphism invariance, the two theories are
also conformal invariant. Now in the Hamiltonian formalism the
constraints $(V,H,S)$ comprise a first-class system. The conformal
constraint generates the following transformations on the phase
space
\ba &&\{h_{ab},S(\lambda)\}=\lambda h_{ab},\quad
\{P^{ab},S(\lambda)\}=-\lambda P^{ab}, \\
&&\{\phi,S(\lambda)\}=-\lambda \phi,\quad \{\pi,S(\lambda)\}=\lambda
\pi. \ea
It is easy to check that the above transformations coincide
with those of spacetime conformal transformations
(\ref{conformalt}). Thus all the physical meaning of constraints has
been cleared. Because of the extra conformal constraint (\ref{sc}),
the physical degrees of freedom of this special kind of STT are
equal to those of GR.

The connection-dynamical formalism for these special theories can
also be obtained by the canonical transformation to the new
variables (\ref{newvaribles}). Then the total Hamiltonian can be
expressed as a linear combination \ba
H_{total}=\ints\Lambda^i\mathcal {G}_i+N^aV_a+NH+\lambda S,\ea where
the Gaussian and diffeomorphism constraints keep the same form as
Eqs. (\ref{GC}) and (\ref{diff}), while the Hamiltonian and the
conformal constraints read respectively
\ba  H
&=&\frac{\phi}{2}\left[F^j_{ab}-(\gamma^2+\frac{1}{\phi^2})\varepsilon_{jmn}\tilde{K}^m_a\tilde{K}^n_b\right]
\frac{\varepsilon_{jkl}
E^a_kE^b_l}{\sqrt{h}}\nn\\
&-&\frac{3}{4\phi}\sqrt{h}(D_a\phi)
D^a\phi+\sqrt{h}D_aD^a\phi+\sqrt{h}\xi(\phi),\label{hamilton1}\\
S&=&\tilde{K}^i_aE^a_i-\pi\phi \label{conformalc}.\ea
The constraints algebra in the connection formalism is closed as \ba
\{\mathcal {G}(\Lambda),H(N)\}&=&0,\\
\{\mathcal {G}(\Lambda),S(\lambda)\}&=&0,\\
\{S(\lambda),H(M)\}&=&H(\frac{\lambda M}{2}),\\
\{H(N),H(M)\}&=&\mathcal {V}(ND^aM-MD^aN)\nn\\
&+&S(\frac{D_a\phi}{\phi}(ND^aM-MD^aN)) \nn\\
&+&\mathcal
{G}\left((N\partial_aM-M\partial_aN)h^{ab}A_b\right)\nn\\
&-&\frac{[E^aD_aN,E^bD_bM]^i}{h}\mathcal {G}_i\nn\\
&-&\gamma^2\frac{[E^aD_a(\phi N),E^bD_b(\phi M)]^i}{h}\mathcal
{G}_i.\ea
It is obvious that the Poisson bracket among the other constraints are also
weakly equal to zero. Since the initial value formalism in this
sector is a delicate issue \cite{So,olmo}, we leave the comparison
between the Hamiltonian formulation and the Lagrangian formulation
for further work.

\section{Loop quantization}

Based on the connection dynamical formalism, the nonperturbative
loop quantization procedure can be straightforwardly extended to the
STT. The kinematical structure of STT is as same as that of $f(\R)$
theories \cite{Zh11,Zh11b}. The kinematical Hilbert space of the
system is a direct product of the Hilbert space of geometry and that
of scalar field, $\hil_\kin:=\hil^\grav_\kin\otimes \hil^\sca_\kin$,
with the orthonormal spin-scalar-network basis
$T_{\alpha,X}(A,\phi)\equiv T_{\alpha}(A)\otimes T_{X}(\phi)$ over
some graph $\alpha\cup X\subset\Sigma$. Here $\alpha$ and $X$
consist of finite number of curves and points respectively in
$\Sigma$. The basic operators are the quantum analogue of holonomies
$h_e(A)=\mathcal {P}\exp\int_eA_a$ of connections, densitized triads
smeared over 2-surfaces $E(S,f):=\int_S\epsilon_{abc}E^a_if^i$,
point holonomis $U_\lambda=\exp(i\lambda\phi(x))$\cite{As03}, and
scalar momenta smeared on 3-dimensional regions $\pi(R):=\int_R
d^3x\pi(x)$. Note that the spatial geometric operator of LQG, such
as the area\cite{Ro95} , the volume\cite{As97} and the
length\cite{Th98,Ma10} operators, are still valid in
$\hil^\grav_\kin$, though their properties in the physical Hilbert
space still need to be clarified \cite{Th09, Rovelli}. As in LQG, it
is straightforward to promote the Gaussian constraint $\mathcal
{G}(\Lambda)$ to a well-defined operator\cite{Th07,Ma07}. It's
kernel is the internal gauge invariant Hilbert space $\mathcal
{H}_G$ with gauge invariant spin-scalar-network basis. Since the
diffeomorphisms of $\Sigma$ act covariantly on the cylindrical
functions in $\mathcal {H}_G$, the so-called group averaging
technique can be employed to solve the diffeomorphism
constraint\cite{As04,Ma07}. Thus we can also obtain the desired
diffeomorphism and gauge invariant Hilbert space $\mathcal
{H}_{Diff}$ for the STT.

\subsection{Sector of $\omega(\phi)\neq -3/2$ }

While the kinematical framework of LQG and polymer-like scalar field
have been straight-forwardly extended to the STT, the nontrivial
task in the case $\omega(\phi)\neq -3/2$ is to implement the
Hamiltonian constraint (\ref{hamilton}) at quantum level. In order
to compare the Hamiltonian constraint of STT with that of metric
$f(\R)$ theories in connection formalism, we write Eq.
(\ref{hamilton}) as $H(N)=\sum^8_{i=1}H_i$ in regular order. It is
easy to see that the terms $H_1,H_2,H_7,H_8$ just keep the same form
as those in $f(\R)$ theories (see Eq.(39) in Ref.\cite{Zh11b}), and
the $H_3,H_4,H_5$ terms are also similar to the corresponding terms
in $f(\R)$ theories. Here the differences are only reflected by the
coefficients as a certain functions of $\phi$. Now we come to the
completely new term, $H_6=\int_\Sigma
d^3xN\frac{\omega(\phi)}{2\phi}\sqrt{h}(D_a\phi) D^a\phi $. This
term is somehow like the kinetic term of a Klein-Gordon field which
was dealt with in Ref.\cite{Ma06}. We can introduce the well-defined
operators $\phi,\phi^{-1}$ as in Ref. \cite{Zh11b}. It is reasonable
to believe that function $\omega(\phi)$ can also be quantized
\cite{Zh11b}. By the same regularization techniques as in
Refs.\cite{Ma06,Zh11b}, we triangulate $\Sigma$ in adaptation to
some graph $\alpha$ underling a cylindrical function in $\hil_\kin$
and reexpress connections by holonomies. The corresponding regulated
operator acts on a basis vector $T_{\alpha,X}$ over some graph
$\alpha\cup X$ as
\ba \hat{H}^\varepsilon_6\cdot T_{\alpha,X}
&=&\lim_{\epsilon\rightarrow 0}\frac{2^{17}N(v)\hat{\omega}(\phi)
}{3^6\gamma^4(i\lambda_0)^2(i\hbar)^4}\hat{\phi}^{-1}(v)
\chi_\epsilon(v-v')
\nn\\
&\times&\sum_{v\in\alpha(v)}\frac{1}{E(v)}\sum_{v(\Delta)=v}\epsilon(s_L
s_M s_N)\epsilon^{LMN}\hat{U}^{-1}_{\lambda_0}(\phi(s_{s_L(\Delta_{v})}))\nn\\
&\times&
[\hat{U}_{\lambda_0}(\phi(t_{s_L(\Delta_{v})}))-\hat{U}_{\lambda_0}(\phi(s_{s_L(\Delta_{v})}))]\nn\\
&\times&\Tr(\tau_i\hat{h}_{s_M(\Delta_{v})}[\hat{h}^{-1}_{s_M(\Delta_{v})},(\hat{V}_{U^\epsilon_{v}})^{3/4}]
\hat{h}_{s_N(\Delta_{v})}[\hat{h}^{-1}_{s_N(\Delta_{v})},(\hat{V}_{U^\epsilon_{v}})^{3/4}]) \nn\\
&\times&\sum_{v'\in\alpha(v)}\frac{1}{E(v')}\sum_{v(\Delta')=v'}\epsilon(s_I
s_J s_K)\epsilon^{IJK}\hat{U}^{-1}_{\lambda_0}(\phi(s_{s_I(\Delta_{v'})}))\nn\\
&\times&
[\hat{U}_{\lambda_0}(\phi(t_{s_I(\Delta_{v'})}))-\hat{U}_{\lambda_0}(\phi(s_{s_I(\Delta_{v'})}))]\nn\\
&\times&\Tr(\tau_i\hat{h}_{s_J(\Delta_{v'})}[\hat{h}^{-1}_{s_J(\Delta_{v'})},(\hat{V}_{U^\epsilon_{v'}})^{3/4}]
\hat{h}_{s_K(\Delta_{v'})}[\hat{h}^{-1}_{s_K(\Delta_{v'})},(\hat{V}_{U^\epsilon_{v'}})^{3/4}])\cdot
T_{\alpha,X}. \label{H6}\ea
We refer to \cite{Zh11b} for the meaning
of notations in Eq.(\ref{H6}). It is easy to see that the action of
$\hat{H}^\varepsilon_6$ on $ T_{\alpha,X}$ is graph changing. It
adds a finite number of vertices at $t(s_I(v))=\varepsilon$ for
edges $e_I(t)$ starting from each high-valent vertex of $\alpha$. As
a result, the family of operators $\hat{H}^\varepsilon_6(N)$ fails
to be weakly convergent when $\varepsilon\rightarrow 0$. However,
due to the diffeomorphism covariant properties of the triangulation,
the limit operator can be well defined via the so-called uniform
Rovelli-Smolin topology induced by diffeomorphism-invariant states
$\Phi_{Diff}$ as:
\ba \Phi_{Diff}(\hat{H}_6\cdot
T_{\alpha,X})=\lim_{\varepsilon\rightarrow
0}(\Phi_{Diff}|\hat{H}^\varepsilon_{6}|T_{\alpha,X}\rangle. \ea
It
is obviously that the limit is independent of $\varepsilon$. Hence
both the regulators $\epsilon$ and $\varepsilon$ can be removed. We
then have
\ba \hat{H}_6\cdot T_{\alpha,X} &=&\sum_{v\in
V(\alpha)}\frac{2^{17}N(v)\hat{\omega}(\phi) }{3^6\gamma^4(i\lambda_0)^2(i\hbar)^4E^2(v)}\hat{\phi}^{-1}(v)\nn\\
&\times&\sum_{v(\Delta)=v(\Delta')=v}\epsilon(s_L
s_M s_N)\epsilon^{LMN}\hat{U}^{-1}_{\lambda_0}(\phi(s_{s_L(\Delta)}))\nn\\
&\times&
[\hat{U}_{\lambda_0}(\phi(t_{s_L(\Delta)}))-\hat{U}_{\lambda_0}(\phi(s_{s_L(\Delta)}))]\nn\\
&\times&\Tr(\tau_i\hat{h}_{s_M(\Delta)}[\hat{h}^{-1}_{s_M(\Delta)},(\hat{V}_v)^{3/4}]
\hat{h}_{s_N(\Delta)}[\hat{h}^{-1}_{s_N(\Delta)},(\hat{V}_v)^{3/4}]) \nn\\
&\times&\epsilon(s_I
s_J s_K)\epsilon^{IJK}\hat{U}^{-1}_{\lambda_0}(\phi(s_{s_I(\Delta')}))\nn\\
&\times&
[\hat{U}_{\lambda_0}(\phi(t_{s_I(\Delta')}))-\hat{U}_{\lambda_0}(\phi(s_{s_I(\Delta')}))]\nn\\
&\times&\Tr(\tau_i\hat{h}_{s_J(\Delta')})[\hat{h}^{-1}_{s_J(\Delta')},(\hat{V}_{v})^{3/4}]
\hat{h}_{s_K(\Delta')}[\hat{h}^{-1}_{s_K(\Delta')},(\hat{V}_{v})^{3/4}])\cdot
T_{\alpha,X} . \ea
In order to simplify the expression, we introduce
$f(\phi)=\frac{1}{3+2\omega(\phi)}$ for the other terms containing
it in $H(N)$, which can also be promoted to a well-defined operator
$\hat{f}(\phi)$. Hence, the terms $H_3,H_4$ and $H_5$ can be
quantized as
\ba \hat{H}_3\cdot T_{\alpha,X} &=&
\sum_{v\in V(\alpha)}\frac{4N(v)\hat{f}(\phi(v))}{\gamma^3(i\hbar)^2}\hat{\phi}^{-1}(v)\nn\\
&\times&
[\hat{\Euc}(1),\sqrt{\hat{V}_v}] [\hat{\Euc}(1),\sqrt{\hat{V}_{v}}]\cdot T_{\alpha,X}, \nn\\
\ea

\ba \hat{H}_4\cdot T_{\alpha,X} &=&-\sum_{v\in V(\alpha)\cap
X}\frac{2^{20}N(v)\hat{f}(\phi(v))}{3^5\gamma^6(i\hbar)^6E^2(v)}\hat{\pi}(v)
\nn\\
&\times&\sum_{v(\Delta)=v(\Delta')=v}\Tr(\tau_i\hat{h}_{s_L(\Delta)}[\hat{h}^{-1}_{s_L(\Delta)},\hat{\kt}])\nn\\
&\times&\epsilon(s_L s_M s_N)\epsilon^{LMN}\nn\\
&\times&\Tr(\tau_i\hat{h}_{s_M(\Delta)}[\hat{h}^{-1}_{s_M(\Delta)},(\hat{V}_{v})^{3/4}]
\hat{h}_{s_N(\Delta)}[\hat{h}^{-1}_{s_N(\Delta)},(\hat{V}_{v})^{3/4}]) \nn\\
&\times&\epsilon(s_I s_J s_K)\epsilon^{IJK}\nn\\
&\times&\Tr(\hat{h}_{s_I(\Delta')}[\hat{h}^{-1}_{s_I(\Delta')},(\hat{V}_{v})^{1/2}]
\hat{h}_{s_J(\Delta')}[\hat{h}^{-1}_{s_J(\Delta')},(\hat{V}_{v})^{1/2}] \nn\\
&\times&\hat{h}_{s_K(\Delta')}[\hat{h}^{-1}_{s_K(\Delta')},(\hat{V}_{v})^{1/2}])\cdot
T_{\alpha,X}, \ea

\ba \hat{H}_5\cdot T_{\alpha,X} &=&\sum_{v\in V(\alpha)\cap
X}\frac{2^{18}N(v)\hat{f}(\phi(v))}{3^4\gamma^6(i\hbar)^6E^2(v)}
\hat{\phi}(v)\hat{\pi}(v)\hat{\pi}(v) \nn\\
&\times&\sum_{v(\Delta)=v(\Delta')=v}\epsilon(s_I s_J
s_K)\epsilon^{IJK}\nn\\
&\times&\Tr(\hat{h}_{s_I(\Delta)}
[\hat{h}^{-1}_{s_I(\Delta)},(\hat{V}_{v})^{1/2}]
\hat{h}_{s_J(\Delta)}[\hat{h}^{-1}_{s_J(\Delta)},(\hat{V}_{v})^{1/2}] \nn\\
&\times&\hat{h}_{s_K(\Delta)}[\hat{h}^{-1}_{s_K(\Delta)},(\hat{V}_{v})^{1/2}]) \nn\\
&\times&\epsilon(s_L s_M s_N)\epsilon^{LMN}\nn\\
&\times&\Tr(\hat{h}_{s_L(\Delta')}[\hat{h}^{-1}_{s_L(\Delta')},(\hat{V}_{v})^{1/2}]
\hat{h}_{s_M(\Delta')}[\hat{h}^{-1}_{s_M(\Delta')},(\hat{V}_{v})^{1/2}] \nn\\
&\times&\hat{h}_{s_N(\Delta')}[\hat{h}^{-1}_{s_N(\Delta')},(\hat{V}_{v})^{1/2}])\cdot
T_{\alpha,X}. \ea
Therefore, the total Hamiltonian constraint in
this sector has been quantized as a well-defined operator
$\hat{H}(N)=\sum_{i=1}^8\hat{H}_i$ in $\hil_\kin$. It is easy to see
that $\hat{H}(N)$ is internal gauge invariant and diffeomorphism
covariant. Hence it is at least well defined in the gauge invariant
Hilbert space $\hil_G$. However, it is difficult to define
$\hat{H}(N)$ directly on $\hil_{Diff}$. Moreover, as in $f(R)$
theories the constraint algebra do not form a Lie algebra. This
might lead to quantum anomaly after quantization.

\subsection{Sector of $\omega(\phi)= -3/2$ }

In the case $\omega(\phi)= -3/2$, there is an extra conformal
constraint (\ref{conformalc}), whose smeared version $S(\lambda)$
has to be promoted as an operator. Note that both $\phi$ and
$\pi(R)$ are already well-defined operators. To quantize the
conformal constraint $S(\lambda)$, we use the classical identity
\ba
\kt\equiv\int_\Sigma
d^3x\tilde{K}^i_aE^a_i=\gamma^{-\frac32}\{\Euc(1),V\}. \ea
Here the
Euclidean scalar constraint $\Euc(1)$ by definition was: \ba
\Euc(1)&=&\frac{1}{2}\int_\Sigma d^3xF^j_{ab}\frac{\varepsilon_{jkl}
E^a_kE^b_l}{\sqrt{h}}. \ea
Both $\Euc$ and the volume $V$ under
consideration have been quantized in LQG. Now it is easy to promote
$S(\lambda)$ as a well-defined operator by acting on a given basis
vector $T_{\alpha,X}\in\hil_\kin$ as
\ba \hat{S}(\lambda)\cdot
T_{\alpha,X} &=& \left(\sum_{v\in
V(\alpha)}\frac{\lambda(v)}{\gamma^{3/2}(i\hbar)}[\hat{H}^E(1),\hat{V}_v]-\sum_{x\in
X}\lambda(x)\hat{\phi}(x)\hat{\pi}(x)\right)\cdot T_{\alpha,X}.\ea
It is easy to see that $\hat{S}(\lambda)$ is internal gauge
invariant, diffeomorphism covariant and graph-changing. Thus it is
well defined in $\hil_G$. Note that the Hamiltonian constraint
operator in this sector is similar to that in the sector of
$\omega(\phi)\neq -3/2$. The difference is that $\omega(\phi)= -3/2$
now. Hence we write Eq. (\ref{hamilton1}) as $H(N)=\sum^5_{i=1}H_i$
in regular order. It is easy to see that the terms $H_1,H_2,H_4,H_5$
just keep the same form as those in $f(\R)$ theories, while the $H_3$ can be quantized as
\ba
\hat{H}_3\cdot T_{\alpha,X} &=&-\sum_{v\in
V(\alpha)}\frac{2^{16}N(v)}{3^5\gamma^4(i\lambda_0)^2(i\hbar)^4E^2(v)}\hat{\phi}^{-1}(v)\nn\\
&\times&\sum_{v(\Delta)=v(\Delta')=v}\epsilon(s_L
s_M s_N)\epsilon^{LMN}\hat{U}^{-1}_{\lambda_0}(\phi(s_{s_L(\Delta)}))\nn\\
&\times&
[\hat{U}_{\lambda_0}(\phi(t_{s_L(\Delta)}))-\hat{U}_{\lambda_0}(\phi(s_{s_L(\Delta)}))]\nn\\
&\times&\Tr(\tau_i\hat{h}_{s_M(\Delta)}[\hat{h}^{-1}_{s_M(\Delta)},(\hat{V}_v)^{3/4}]
\hat{h}_{s_N(\Delta)}[\hat{h}^{-1}_{s_N(\Delta)},(\hat{V}_v)^{3/4}]) \nn\\
&\times&\epsilon(s_I
s_J s_K)\epsilon^{IJK}\hat{U}^{-1}_{\lambda_0}(\phi(s_{s_I(\Delta')}))\nn\\
&\times&
[\hat{U}_{\lambda_0}(\phi(t_{s_I(\Delta')}))-\hat{U}_{\lambda_0}(\phi(s_{s_I(\Delta')}))]\nn\\
&\times&\Tr(\tau_i\hat{h}_{s_J(\Delta')})[\hat{h}^{-1}_{s_J(\Delta')},(\hat{V}_{v})^{3/4}]
\hat{h}_{s_K(\Delta')}[\hat{h}^{-1}_{s_K(\Delta')},(\hat{V}_{v})^{3/4}])\cdot
T_{\alpha,X} . \ea
Thus the total Hamiltonian constraint operator
$\hat{H}(N)=\sum^5_{i=1}\hat{H}_i$ is also well defined in $\hil_G$.

\section{master constraint}

In order to avoid possible quantum anomaly and find the physical
Hilbert space, master constraint programme was first introduced into
LQG by Thiemann in \cite{Th06}. This programme can also be applied
to the above quantum STT.

\subsection{Sector of $\omega(\phi)\neq -3/2$ }
In the case $\omega(\phi)\neq -3/2$, we can employ the master
constraint to implement the Hamiltonian constraint. By definition
the master constraint of the STT classically reads
\ba \mathcal {M}:=\frac12\int_\Sigma
d^3x\frac{\abs{H(x)}^2}{\sqrt{h}}, \label{mcs}\ea
where the
Hamiltonian constraint $H(x)$ is given by Eq. (\ref{hamilton}). The
master constraint can be regulated via a point-splitting strategy
\cite{Ma061} as:
\ba \mathcal {M}^\epsilon=\frac12\int_\Sigma d^3y\int_\Sigma
d^3x\chi_\epsilon(x-y)\frac{{H(x)}}{\sqrt{V_{U^\epsilon_x}}}\frac{{H(y)}}{\sqrt{V_{U^\epsilon_y}}}.
\ea
Introducing a partition $\mathcal {P}$ of the 3-manifold
$\Sigma$ into cells $C$, we have an operator
$\hat{H}^\varepsilon_{C,\beta}$ acting on the internal
gauge-invariant spin-scalar-network basis $T_{s,c}$ in $\hil_G$ via
a state-dependent triangulation,
\ba
\hat{H}^\varepsilon_{C,\alpha}\cdot T_{s,c}=\sum_{v\in
V(\alpha)}\chi_C(v)\hat{H}^\varepsilon_v \cdot
T_{s,c}\label{master1}\ea
where $\alpha$ denotes the underlying
graph of the spin-network state $T_{s}$, $\chi_C$ is the characteristic function over $C$, and
\ba\hat{H}^\varepsilon_v=\sum_{v(\Delta)=v}\hat{H}^{\varepsilon,\Delta}_{GR,v}+\sum^8_{i=3}
\hat{H}^\varepsilon_{i,v}, \ea
with
\ba \hat{H}_{3,v}^\varepsilon &=&
\frac{16\hat{f}(\phi(v))}{\gamma^3(i\hbar)^2}\hat{\phi}^{-1}(v)\nn\\
&\times&
[\hat{\Euc}(1),(\hat{V}_{U^\epsilon_{v}})^{1/4}][\hat{\Euc}(1),(\hat{V}_{U^\epsilon_{v}})^{1/4}],
\ea

\ba \hat{H}^\varepsilon_{4,v}
&=&-\sum_{v(\Delta)=v(\Delta')=v(X)=v}\frac{2^{18}\hat{f}(\phi(v))}{3^3\gamma^6(i\hbar)^6E^2(v)}\hat{\pi}(v)
\nn\\
&\times&\Tr(\tau_i\hat{h}_{s_L(\Delta)}[\hat{h}^{-1}_{s_L(\Delta)},\hat{\kt}])\nn\\
&\times&\epsilon(s_L s_M s_N)\epsilon^{LMN}\Tr(\tau_i\hat{h}_{s_M(\Delta)}[\hat{h}^{-1}_{s_M(\Delta)},(\hat{V}_{U^\epsilon_{v}})^{1/2}]\nn\\
&\times&\hat{h}_{s_N(\Delta)}[\hat{h}^{-1}_{s_N(\Delta)},(\hat{V}_{U^\epsilon_{v}})^{1/2}]) \nn\\
&\times&\epsilon(s_I
s_J s_K)\epsilon^{IJK}\Tr(\hat{h}_{s_I(\Delta')}[\hat{h}^{-1}_{s_I(\Delta')},(\hat{V}_{U^\epsilon_{v}})^{1/2}]\nn\\
&\times&
\hat{h}_{s_J(\Delta')}[\hat{h}^{-1}_{s_J(\Delta')},(\hat{V}_{U^\epsilon_{v}})^{1/2}] \nn\\
&\times&\hat{h}_{s_K(\Delta')}[\hat{h}^{-1}_{s_K(\Delta')},(\hat{V}_{U^\epsilon_{v}})^{1/2}]),\ea

\ba \hat{H}^\varepsilon_{5,v}
&=&\sum_{v(\Delta)=v(\Delta')=v(X)=v}\frac{2^{20}\hat{f}(\phi(v))}{3^4\gamma^6(i\hbar)^6E^2(v)}
\hat{\phi}(v)\hat{\pi}(v)\hat{\pi}(v) \nn\\
&\times&\epsilon(s_I s_J s_K)\epsilon^{IJK}\Tr(\hat{h}_{s_I(\Delta)}
[\hat{h}^{-1}_{s_I(\Delta)},(\hat{V}_{U^\epsilon_{v}})^{1/4}]\nn\\
&\times&
\hat{h}_{s_J(\Delta)}[\hat{h}^{-1}_{s_J(\Delta)},(\hat{V}_{U^\epsilon_{v}})^{1/2}] \nn\\
&\times&\hat{h}_{s_K(\Delta)}[\hat{h}^{-1}_{s_K(\Delta)},(\hat{V}_{U^\epsilon_{v}})^{1/2}]) \nn\\
&\times&\epsilon(s_L s_M s_N)\epsilon^{LMN}\Tr(\hat{h}_{s_L(\Delta')}[\hat{h}^{-1}_{s_L(\Delta')},(\hat{V}_{U^\epsilon_{v}})^{1/4}]\nn\\
&\times&
\hat{h}_{s_M(\Delta')}[\hat{h}^{-1}_{s_M(\Delta')},(\hat{V}_{U^\epsilon_{v}})^{1/2}] \nn\\
&\times&\hat{h}_{s_N(\Delta')}[\hat{h}^{-1}_{s_N(\Delta')},(\hat{V}_{U^\epsilon_{v}})^{1/2}]),
\ea

\ba \hat{H}^\varepsilon_{6,v}
&=&\sum_{v(\Delta)=v(\Delta')=v}\frac{2^{15}\hat{\omega}(\phi)}{3^4\gamma^4(i\lambda_0)^2(i\hbar)^4E^2(v)}\hat{\phi}^{-1}(v)\nn\\
&\times&\epsilon(s_L
s_M s_N)\epsilon^{LMN}\hat{U}^{-1}_{\lambda_0}(\phi(s_{s_L(\Delta)}))\nn\\
&\times&
[\hat{U}_{\lambda_0}(\phi(t_{s_L(\Delta)}))-\hat{U}_{\lambda_0}(\phi(s_{s_L(\Delta)}))]\nn\\
&\times&\Tr(\tau_i\hat{h}_{s_M(\Delta)}[\hat{h}^{-1}_{s_M(\Delta)},(\hat{V}_v)^{1/2}]
\hat{h}_{s_N(\Delta)}[\hat{h}^{-1}_{s_N(\Delta)},(\hat{V}_v)^{3/4}]) \nn\\
&\times&\epsilon(s_I
s_J s_K)\epsilon^{IJK}\hat{U}^{-1}_{\lambda_0}(\phi(s_{s_I(\Delta')}))\nn\\
&\times&
[\hat{U}_{\lambda_0}(\phi(t_{s_I(\Delta')}))-\hat{U}_{\lambda_0}(\phi(s_{s_I(\Delta')}))]\nn\\
&\times&\Tr(\tau_i\hat{h}_{s_J(\Delta')})[\hat{h}^{-1}_{s_J(\Delta')},(\hat{V}_{v})^{1/2}]
\hat{h}_{s_K(\Delta')}[\hat{h}^{-1}_{s_K(\Delta')},(\hat{V}_{v})^{3/4}])
\ea
Note that $\hat{H}^{\varepsilon,\Delta}_{GR,v}$ and
$H^\varepsilon_{7,v}$ keep the same form as the corresponding terms
in \cite{Zh11b}, while $H^\varepsilon_{8,v}$ is twice as the
corresponding term in \cite{Zh11b}. Since the family of operators
$\hat{H}^\varepsilon_{C,\alpha}$ are cylindrically consistent up to
diffeomorphism, the inductive limit operator $\hat{H}_{C}$ is
densely defined in $\hil_G$ by the uniform Rovelli- Smolin topology.
Then we could define master constraint operator $\hat{\mathcal {M}}$
on diffeomorphism invariant states as
\ba( \hat{\mathcal {M}}\Phi_{Diff})T_{s,c}=\lim_{\mathcal
{P}\rightarrow\Sigma,\varepsilon,\varepsilon'\rightarrow
0}\Phi_{Diff}[\frac12\sum_{c\in\mathcal
{P}}\hat{H}^\varepsilon_C(\hat{H}^{\varepsilon'}_C)^\dagger T_{s,c}
]. \ea
Note that our construction of $\hat{\mathcal {M}}$ is
qualitatively similar to those in \cite{Zh11b,Ma06}, although the
quantitative actions are different. Similar to those in
\cite{Ma06,Zh11b}, we can prove that $\hat{\mathcal {M}}$ is a
positive and symmetric operator in $\hil_{Diff}$ and hence admits a
unique self-adjoint Friedrichs extension. It is then possible to
obtain the physical Hilbert space of the quantum STT in this sector
by the direct integral decomposition of $\hil_{Diff}$ with respect
to $\hat{\mathcal {M}}$.

\subsection{Sector of $\omega(\phi)= -3/2$ }
In the case $\omega(\phi)= -3/2$, we need to employ the master
constraint to implement both the Hamiltonian constraint and
conformal constraint. Hence, we define the master constraint for
this sector as
\ba \mathcal {M}:=\frac12\int_\Sigma
d^3x\frac{\abs{H(x)}^2+\abs{S(x)}^2}{\sqrt{h}}, \label{mcs1}\ea
where the Hamiltonian constraint $H(x)$ and the conformal constraint
$S(x)$ are given by Eqs. (\ref{hamilton1}) and (\ref{conformalc})
respectively. It is obvious that \ba \mathcal {M}=0 \Leftrightarrow
H(N)=0 \quad and\quad S(\lambda)=0 \quad\forall N(x),\lambda(x). \ea
Now the constraints also form a Lie algebra. The master constraint
can be regulated via a point-splitting strategy as:
\ba \mathcal {M}^\epsilon=\frac12\int_\Sigma d^3y\int_\Sigma
d^3x\chi_\epsilon(x-y)\frac{H(x)H(y)+S(x)S(y)}{\sqrt{V_{U^\epsilon_x}}\sqrt{V_{U^\epsilon_y}}}.
\ea
Introducing a partition $\mathcal {P}$ of the 3-manifold
$\Sigma$ into cells $C$, we have an operator
$\hat{H}^\varepsilon_{C,\beta}$ acting on spin-scalar-network basis
$T_{s,c}$ in $\hil_G$ via a state-dependent triangulation as Eq.
(\ref{master1}). Here, $\hat{H}^\varepsilon_v$ has less terms than
in Eq. (\ref{master1}) as
\ba\hat{H}^\varepsilon_v=\sum_{v(\Delta)=v}\hat{H}^{\varepsilon,\Delta}_{GR,v}+\sum^5_{i=3}
\hat{H}^\varepsilon_{i,v}, \ea
where
\ba \hat{H}^\varepsilon_{3,v}
&=&-\sum_{v(\Delta)=v(\Delta')=v}\frac{2^{14}}{3^3\gamma^4(i\lambda_0)^2(i\hbar)^4E^2(v)}\hat{\phi}^{-1}(v)\nn\\
&\times&\epsilon(s_L
s_M s_N)\epsilon^{LMN}\hat{U}^{-1}_{\lambda_0}(\phi(s_{s_L(\Delta)}))\nn\\
&\times&
[\hat{U}_{\lambda_0}(\phi(t_{s_L(\Delta)}))-\hat{U}_{\lambda_0}(\phi(s_{s_L(\Delta)}))]\nn\\
&\times&\Tr(\tau_i\hat{h}_{s_M(\Delta)}[\hat{h}^{-1}_{s_M(\Delta)},(\hat{V}_v)^{1/2}]
\hat{h}_{s_N(\Delta)}[\hat{h}^{-1}_{s_N(\Delta)},(\hat{V}_v)^{3/4}]) \nn\\
&\times&\epsilon(s_I
s_J s_K)\epsilon^{IJK}\hat{U}^{-1}_{\lambda_0}(\phi(s_{s_I(\Delta')}))\nn\\
&\times&
[\hat{U}_{\lambda_0}(\phi(t_{s_I(\Delta')}))-\hat{U}_{\lambda_0}(\phi(s_{s_I(\Delta')}))]\nn\\
&\times&\Tr(\tau_i\hat{h}_{s_J(\Delta')})[\hat{h}^{-1}_{s_J(\Delta')},(\hat{V}_{v})^{1/2}]
\hat{h}_{s_K(\Delta')}[\hat{h}^{-1}_{s_K(\Delta')},(\hat{V}_{v})^{3/4}]),
\ea
and $H^\varepsilon_{4,v}$ keep the same form as the
corresponding terms in \cite{Zh11b}, while $H^\varepsilon_{5,v}$ is
twice as the corresponding term in \cite{Zh11b}. Similarly, the
operator corresponding to the conformal constraint can be defined as
\ba \hat{S}^\varepsilon_{C,\alpha}\cdot T_{s,c}=\sum_{v\in
V(\alpha)}\chi_C(v)\hat{S}^\varepsilon_v \cdot T_{s,c}\ea where
\ba\hat{S}^\varepsilon_v=\hat{S}^\varepsilon_{1,v}+\hat{S}^\varepsilon_{2,v},
\ea with \ba \hat{S}_{1,v}^\varepsilon &=&
\frac{2}{\gamma^{3/2}(i\hbar)}
[\hat{H}^E(1),(\hat{V}_{U^\epsilon_{v}})^{1/2}], \ea

\ba \hat{S}^\varepsilon_{2,v}
&=&-\sum_{v(\Delta)=v(X)=v}\frac{2^{7}}{3\gamma^3(i\hbar)^3E(v)}
\hat{\phi}(v)\hat{\pi}(v) \nn\\
&\times&\epsilon(s_I s_J s_K)\epsilon^{IJK}\Tr(\hat{h}_{s_I(\Delta)}
[\hat{h}^{-1}_{s_I(\Delta)},(\hat{V}_{U^\epsilon_{v}})^{1/2}]\nn\\
&\times&
\hat{h}_{s_J(\Delta)}[\hat{h}^{-1}_{s_J(\Delta)},(\hat{V}_{U^\epsilon_{v}})^{1/2}] \nn\\
&\times&\hat{h}_{s_K(\Delta)}[\hat{h}^{-1}_{s_K(\Delta)},(\hat{V}_{U^\epsilon_{v}})^{1/2}]).
\ea
Since the family of operators $\hat{H}^\varepsilon_{C,\alpha}$
and $\hat{S}^\varepsilon_{C,\alpha}$ are cylindrically consistent up
to diffeomorphism, the inductive limit operator $\hat{H}_{C}$ and
$\hat{S}_{C}$ are densely defined in $\hil_G$ by the uniform
Rovelli- Smolin topology. Then we could define master constraint
operator $\hat{\mathcal {M}}$ on diffeomorphism invariant states as
\ba( \hat{\mathcal {M}}\Phi_{Diff})T_{s,c}=\lim_{\mathcal
{P}\rightarrow\Sigma,\varepsilon,\varepsilon'\rightarrow
0}\Phi_{Diff}[\frac12\sum_{c\in\mathcal
{P}}\left(\hat{H}^\varepsilon_C(\hat{H}^{\varepsilon'}_C)^\dagger+\hat{S}^\varepsilon_C(\hat{S}^{\varepsilon'}_C)^\dagger\right)
T_{s,c} ]. \ea
Similar to those in \cite{Ma06,Zh11b}, we can prove
that $\hat{\mathcal {M}}$ is a positive and symmetric operator in
$\hil_{Diff}$ and hence admits a unique self-adjoint Friedrichs
extension. It is then also possible to obtain the physical Hilbert
space of the quantum STT in this sector by the direct integral
decomposition of $\hil_{Diff}$ with respect to $\hat{\mathcal {M}}$.

\section{Concluding Remarks}

STT have received increased attention in issues of "dark Universe"
and nontrivial tests on gravity beyond GR. These kinds of theories
have also become popular in unification schemes such as string
theory. Hence it is desirable to study the Hamiltonian formulation
of general STT. The first achievement in this paper is the detailed Hamiltonian structure and connection dynamics of STT. By doing Hamiltonian analysis, we have derived the
Hamiltonian formulation of STT of gravity from their Lagrangian
formulation. Two sectors of STT are marked off by the coupling
parameter $\omega(\phi)$. In the sector of $\omega(\phi)\neq -3/2$,
the canonical structure and constraint algebra of STT are similar to
those of GR coupled with a scalar field. In the sector of
$\omega(\phi)= -3/2$, the feasible theories are restricted
and a new primary constraint generating conformal transformations of
spacetime is obtained. The canonical structure and constraint algebra
are also obtained. Note that Palatini $f(\R)$ theories are
equivalent to this sector of STT. The successful background independent LQG
relies on the key observation that GR can be cast into the
connection dynamics with structure group of $SU(2)$. We have shown
that the connection dynamical formalism
of the STT can also be obtained by canonical transformations from
the geometric dynamics.

The second achievement of this paper is the nonperturbative loop quantization of STT. Based on the $su(2)$-connection dynamical
formalism, LQG has been successfully extended to the STT by coupling
to a polymer-like scalar field. The quantum kinematical structure of
STT is as same as that of loop quantum $f(\R)$ theories. Thus the
important physical result that both the area and the volume are
discrete at kinematic level remains valid for quantum STT of
gravity. While the dynamics of STT is more general than that of
$f(\R)$ theories, the Hamiltonian constraint operator and
master constraint operator for STT can also be well defined in both sectors. In particular, in the sector $\omega(\phi)= -3/2$, the conformal constraint can also be quantized as a well-defined operator. Hence
the classical STT in both sectors have been successfully quantized
non-perturbatively. This guarantees the existence of the STT of
gravity at fundamental quantum level. Meanwhile, besides GR and metric
$f(\R)$ theories, LQG method is also valid for general
STT of gravity.

It should be noted that there are still many aspects of the
connection formalism and loop quantization of STT which deserve
discovering. For examples, it is still desirable to derive the
connection dynamics of STT by variational principle. The
semiclassical analysis of loop quantum STT is yet to be done. To
further explore the physical contents of the loop quantum STT, we
would like to study its applications to cosmology and black holes in
future works. Moreover, one would also like to quantize STT by the
covariant spin foam approach.

\begin{acknowledgements}

This work is supported by NSFC (Grant No.10975017) and the
Fundamental Research Funds for the Central Universities.
\end{acknowledgements}

\section*{Appendix A}

We first use $(\kt^i_a,E^b_j)$ and $(\phi,\pi)$ as canonical variables to
derive the Poisson bracket (\ref{eqsb}) of STT with $\omega(\phi)\neq -3/2$.
The Hamiltonian constraint (\ref{hamilton}) can also be
written as
\ba H
&=&\frac{1}{2\sqrt{h}\phi}(\kt^i_aE^b_i\kt^j_bE^a_j-\kt^i_aE^a_i\kt^j_bE^b_j)
-\frac12\phi\sqrt{h}R+\sqrt{h}\xi(\phi)\nn\\
&+&\frac{1}{(3+2\omega)\phi\sqrt{h}}(\kt^i_aE^a_i+\pi\phi)^2+\frac{\omega}{2\phi}\sqrt{h}D_a\phi
D^a\phi+\sqrt{h}D_aD^a\phi.\ea
To calculate the Poisson bracket
between two smeared Hamiltonian constraints, we notice that the
non-vanishing contributions come only form the terms which contain
the derivative of canonical variables. Those terms are $\ints
\frac{N\sqrt{h}\omega}{2\phi}D_a\phi D^a\phi$, which only contains
 the derivative of $\phi$, and $\ints
d^3xN\sqrt{h}D_aD^a\phi$, which contains both the derivative of
$E^b_j$ and the derivative of $\phi$, and $\ints-\frac12\phi
N\sqrt{h}R$, which only contain the derivative of $E^b_j$. Hence we
first use $\{\phi(x),\pi(y)\}=\delta^3(x,y)$ to calculate
\ba
&\{&\ints
N\sqrt{h}D_aD^a\phi,\ints\frac{M}{(3+2\omega)\phi\sqrt{h}}(\kt^i_aE^a_i+\pi\phi)^2\}_{(\phi,\pi)}-M
\leftrightarrow N \nn\\
&=&\ints(MD_aD^aN-ND_aD^aM)\frac{2}{(3+2\omega)}(\kt^i_aE^a_i+\pi\phi)\nn\\
&=&\frac{2}{(3+2\omega)}\ints(ND^aM-MD^aN)D_a(\pi\phi+\kt^i_bE^b_i),\label{hstart}\ea
and
\ba &\{&\ints \frac{N\sqrt{h}\omega}{2\phi}D_a\phi
D^a\phi,\ints\frac{M}{(3+2\omega)\phi\sqrt{h}}(\kt^i_bE^b_i+\pi\phi)^2\}_{(\phi,\pi)}-M
\leftrightarrow N \nn\\
&=&-\ints(MD^aN-ND^aM)\frac{\omega D_a\phi}{\phi}\frac{2}{(3+2\omega)}(\kt^i_bE^b_i+\pi\phi)\nn\\
&=&\frac{2\omega}{(3+2\omega)}\ints(ND^aM-MD^aN)(\pi\phi+\kt^i_bE^b_i)\frac{D_a\phi}{\phi}.\ea
Note also that
\ba
N\sqrt{h}D_aD^a\phi=N\sqrt{h}h^{ab}(\partial_a\partial_b\phi-\Gamma^c_{ab}\partial_c\phi).
\ea
Since only  $\Gamma^c_{ab}$ contains the derivative of $E^a_i$
in above equation, we consider
\ba
N\sqrt{h}h^{ab}\Gamma^c_{ab}\partial_c\phi&=&\frac
N2\sqrt{h}h^{ab}(\partial_c\phi)(h^{cd}(-\partial_ah_{bd}-\partial_bh_{ad}+\partial_dh_{ab}))\nn\\
&=&\frac
N2\sqrt{h}(\partial_c\phi)(2\partial_ah^{ac}-h_{ab}\partial^ch^{ab})\nn\\
&=&\frac
N2\sqrt{h}(\partial_c\phi)(2\partial_a(\frac{E^a_iE^c_i}{h})-h_{ab}\partial^c(\frac{E^a_iE^b_i}{h})).
\ea
Therefore, we use
$\{\tilde{K}^j_a(x),E_k^b(y)\}=\delta^b_a\delta^j_k\delta(x,y)$ to
calculate \ba &\{&\ints
N\sqrt{h}(\partial_c\phi)\partial_a(\frac{E^a_iE^c_i}{h}),\ints\frac{M}{2\sqrt{h}}
(\frac1\phi(\kt^l_dE^b_l\kt^j_bE^d_j-\frac{1+2\omega}{3+2\omega}\kt^l_dE^d_l\kt^j_bE^b_j)
+\frac4{3+2\omega}\kt^l_dE^d_l\pi)\}_{(\tilde{K},E)}-M
\leftrightarrow N \nn\\
&=&\ints\frac12M(\partial_aN)(D_c\phi)\frac{2E^c_i}{h}(\frac2\phi(E^b_i\kt^j_bE^a_j-\frac{1+2\omega}{3+2\omega}E^a_i\kt^j_bE^b_j)
+\frac4{3+2\omega}E^a_i\pi))\nn\\
&+&\frac12M(\partial_aN)(D_c\phi)\frac{E^a_iE^c_i}{h}(-E^j_d)(\frac2\phi(E^b_j\kt^m_bE^d_m-\frac{1+2\omega}{3+2\omega}E^d_j\kt^m_bE^b_m)
+\frac4{3+2\omega}E^d_j\pi))-M
\leftrightarrow N \nn\\
\ea and \ba &\{& \ints-\frac N2\sqrt{h}(\partial_c\phi)
h_{ae}\partial^c(\frac{E^a_iE^e_i}{h}),\ints\frac{M}{2\sqrt{h}}
(\frac1\phi(\kt^l_dE^b_l\kt^j_bE^d_j-\frac{1+2\omega}{3+2\omega}\kt^l_dE^d_l\kt^j_bE^b_j)+\frac4{3+2\omega}\kt^l_dE^d_l\pi)\}_{(\tilde{K},E)}-M
\leftrightarrow N \nn\\
&=&\ints-\frac14M\partial^cND_c\phi
h_{ae}\frac{2E^e_i}{h}(\frac2\phi(E^b_i\kt^j_bE^a_j-\frac{1+2\omega}{3+2\omega}E^a_i\kt^j_bE^b_j)
+\frac4{3+2\omega}E^a_i\pi))\nn\\
&-&\frac14M\partial_aND_c\phi\frac{E^a_iE^c_i}{h}(-3E^j_d)(\frac2\phi(E^b_j\kt^m_bE^d_m-\frac{1+2\omega}{3+2\omega}E^d_j\kt^m_bE^b_m)
+\frac4{3+2\omega}E^d_j\pi))-M
\leftrightarrow N. \nn\\
\ea
The combination of above two Poisson brackets equals to
\ba
\ints(ND^aM-MD^aN)(-\frac1{3+2\omega}\pi
D_a\phi-\frac2\phi(\kt^j_bE^c_jh_{ac}D^b\phi-\frac{1+\omega}{3+2\omega}\kt^j_bE^b_jD_a\phi)).
\ea
The variation of the terms containing a derivative  in
$\ints-\frac12\phi N\sqrt{h}R$ reads \ba
&&\ints\frac12\sqrt{h}(-D^aD^b(\phi N)+h^{ab}D_cD^c(\phi N))\delta
h_{ab}\nn\\
&=&\ints\frac12\sqrt{h}(D_aD_b(\phi N)-h_{ab}D_cD^c(\phi
N))\delta h^{ab}\nn\\
&=&\ints\frac12\sqrt{h}(D_aD_b(\phi N)-h_{ab}D_cD^c(\phi N))\delta
(\frac{E^a_iE^b_i}{h}).\ea Thus we have \ba &\{&\ints-\frac12\phi
N\sqrt{h}R,\ints\frac{M}{2\sqrt{h}}(\frac1\phi(\kt^l_dE^e_l\kt^j_eE^d_j-\frac{1+2\omega}{3+2\omega}\kt^j_dE^d_j\kt^m_eE^e_m)
+\frac4{3+2\omega}\kt^l_dE^d_l\pi)\}
-M \leftrightarrow N \nn\\
&=&\ints-\frac14(MD_aD_b(\phi N)-h_{ab}MD_cD^c(\phi
N))\frac{2E^b_i}{h}(\frac2\phi(E^e_i\kt^j_eE^a_j-\frac{1+2\omega}{3+2\omega}\kt^j_dE^d_jE^a_i)+\frac4{3+2\omega}E^a_i\pi)\nn\\
&-&\frac14(-2MD_cD^c(\phi
N))(-E^i_a)(\frac2\phi(E^e_i\kt^j_eE^a_j-\frac{1+2\omega}{3+2\omega}\kt^j_dE^d_jE^a_i)+\frac4{3+2\omega}E^a_i\pi)-M
\leftrightarrow N \nn\\
&=&\ints-(MD_aD_b(\phi N)-h_{ab}MD_cD^c(\phi
N))h^{be}\frac1\phi\kt^j_eE^a_j\nn\\
&-&MD_cD^c(\phi
N)(\frac{2}{\phi(3+2\omega)}\kt^j_dE^d_j+\frac{2}{3+2\omega}\pi) -M
\leftrightarrow N \nn\\
&=&\ints-MD_aD^b(\phi N)\frac1\phi\kt^j_bE^a_j +MD_cD^c(\phi
N)(\frac{1+2\omega}{\phi(3+2\omega)}\kt^j_dE^d_j-\frac2{3+2\omega}\pi)
-M
\leftrightarrow N \nn\\
&=&\ints(ND_aD^b(\phi M)-MD_aD^b(\phi N))\frac1\phi\kt^j_bE^a_j+
(ND_cD^c(\phi M)-MD_cD^c(\phi
N))(\frac2{3+2\omega}\pi-\frac{1+2\omega}{\phi(3+2\omega)}\kt^j_dE^d_j)\nn\\
&=&\ints(ND_cD^cM-MD_cD^cN)(\frac2{3+2\omega}\pi\phi-\frac{1+2\omega}{3+2\omega}\kt^j_aE^a_j)
+(ND_cM-MD_cN)D^c\phi(\frac4{3+2\omega}\pi-\frac{2(1+2\omega)}{\phi(3+2\omega)}\kt^j_aE^a_j)\nn\\
&+&(ND_aD^bM-MD_aD^bN)\kt^j_bE^a_j+(ND_aM-MD_aN)\frac{2D^b\phi}{\phi}\kt^j_bE^a_j.\label{hend1}
\ea Taking account of Eqs.(\ref{hstart})-(\ref{hend1}), we obtain
\ba&&\{H(N),H(M)\}=\nn\\
&&\ints(ND_cD^cM-MD_cD^cN)(-\kt^j_aE^a_j)+(ND^aM-MD^aN)(\pi
D_a\phi)\nn\\
&+&(ND_aD^bM-MD_aD^bN)\kt^j_bE^a_j \nn\\
&=&\ints(ND^aM-MD^aN)(D_a(\kt^j_cE^c_j)-D_b(\kt^j_aE^b_j)+\pi
D_a\phi)-(D_aMD^bN-D^bMD_aN)\kt^j_bE^a_j\nn\\
&=&\ints(ND^aM-MD^aN)V_a+\frac{[E^aD_aN,E^bD_bM]^i}{h}\mathcal
{G}_i,\ea where we used the following identity \ba
-[(D_aM)D^bN-(D^bM)D_aN]\kt^j_bE^a_j&=&-((D_aM)D_cN-(D_cM)D_aN)h^{b
c}E^a_j\kt^j_b\nn\\
&=&-2(D_{[a}M)(D_{c]}N)\frac{E^b_iE^c_i}{h}E^a_j\kt^j_b \nn\\
&=&-2(D_{a}M)(D_{c}N)\frac{E^{[a}_jE^{c]}_i}{h}\kt^j_bE^{ib}\nn\\
&=&-\epsilon^{ijk}(D_{a}M)(D_{c}N)\frac{E^{a}_{j}E^{c}_{i}}{h}\kt^m_bE^{nb}\varepsilon_{kmn}\nn\\
&=&-\frac{[E^aD_aN,E^bD_bM]^k}{h}\mathcal {G}_k.\ea
Using above
result and shift conjugate pair $(\kt^i_a,E^b_j)$ to
$(A^i_a,E^b_j)$, we can easily get the Poisson bracket (\ref{eqsb})
between the smeared Hamiltonian constraints.

Now we use
$(h_{ab},p^{cd})$ and $(\phi,\pi)$ as canonical variables to derive
the
Poisson brackets (\ref{HH}) and (\ref{Sc}) in the sector $\omega(\phi)= -3/2$. The Hamiltonian constraint and conformal
constraint can be read from Eqs. (\ref{hc1}) and (\ref{sc})
respectively. To calculate the Poisson bracket between two smeared
Hamiltonian constraints by the symplectic structure
(\ref{poission}), we notice that the non-vanishing contributions
come only form the terms which contain the derivative of canonical
variables. Those terms are $\ints d^3xN\sqrt{h}D_aD^a\phi$, which
contains both the derivative of $h_{ab}$ and the derivative of
$\phi$, and $\ints-\frac12\phi N\sqrt{h}R$, which only contain the
derivative of $h_{ab}$. Hence we first notice that
\ba
N\sqrt{h}D_aD^a\phi=N\sqrt{h}h^{ab}(\partial_a\partial_b\phi-\Gamma^c_{ab}\partial_c\phi).
\ea
Since only  $\Gamma^c_{ab}$ contains the derivative of $h_{ab}$
in above equation, we consider \ba
&&-N\sqrt{h}h^{ab}\Gamma^c_{ab}\partial_c\phi\nn\\
&=&\frac
N2\sqrt{h}h^{ab}(\partial_c\phi)(h^{cd}(-\partial_ah_{bd}-\partial_bh_{ad}+\partial_dh_{ab}))\nn\\
&=&\frac
N2\sqrt{h}(\partial_c\phi)(2\partial_ah^{ac}-h_{ab}\partial^ch^{ab}).\nn\\
\ea Thus we calculate \ba &&\{\ints
N\sqrt{h}(\partial_c\phi)\partial_a(h^{ac}),\ints\frac{2M}{\sqrt{h}}
(\frac{p_{bd}p^{bd}-\frac12p^2}{\phi})\}_{(h,p)}-M
\leftrightarrow N \nn\\
&=&\ints2(M\partial_aN-N\partial_aM)(D_c\phi)(\frac{2(p^{ac}-\frac12ph^{ac})}{\phi}),\nn\\
\ea and \ba &&\{ \ints-\frac N2\sqrt{h}(\partial_c\phi)
h_{ae}\partial^c(h^{ae}),\ints\frac{2M}{\sqrt{h}}
(\frac{p_{bd}p^{bd}-\frac12p^2}{\phi})\}_{(h,p)}-M
\leftrightarrow N \nn\\
&=&\ints(M\partial^c N-N\partial^c M)(D_c\phi)\frac{p}{\phi}. \ea
The combination of above two Poisson brackets gives \ba &&\{\ints
N\sqrt{h}D_aD^a\phi,\ints\frac{2M}{\sqrt{h}}
(\frac{p_{cd}p^{cd}-\frac12p^2}{\phi})\}_{(h,p)}-M
\leftrightarrow N \nn\\
&=&\ints(M\partial_aN-N\partial_aM)(D_c\phi)(\frac{4(p^{ac}-\frac14ph^{ac})}{\phi}).\label{h2}
\ea
On the other hand, the variation of the terms containing the
derivative in $\ints-\frac12\phi N\sqrt{h}R$ reads
\ba
&&\ints\frac12\sqrt{h}(-D^aD^b(\phi N)+h^{ab}D_cD^c(\phi N))\delta
h_{ab}\nn\\
&=&\ints\frac12\sqrt{h}(D_aD_b(\phi N)-h_{ab}D_cD^c(\phi N))\delta
h^{ab}.\nn\ea
Thus we have \ba &\{&\ints-\frac12\phi
N\sqrt{h}R,\ints\frac{2M}{\sqrt{h}}
(\frac{p_{ab}p^{ab}-\frac12p^2}{\phi})\}
-M \leftrightarrow N \nn\\
&=&-\ints(MD_aD_b(\phi N)-h_{ab}MD_cD^c(\phi
N))(\frac{2(p^{ab}-\frac12ph^{ab})}{\phi})-M
\leftrightarrow N \nn\\
&=&\ints(ND^aM-MD^aN)(-2D^bp_{ab}+4(D^b\phi)\frac{p_{ab}}{\phi}).\label{hend}
\ea
Taking account of Eqs. (\ref{h2}) and (\ref{hend}), we obtain
\ba \{H(N),H(M)\}&=&\ints(ND^aM-MD^aN)(-2D^bp_{ab}+\pi
D_a\phi+\frac{D_a\phi}{\phi}(p-\phi\pi))\nn\\
&=&V(ND^aM-MD^aN)+S(\frac{D_a\phi}{\phi}(ND^aM-MD^aN)). \ea
Now, we
calculate the Poisson bracket between conformal constraint and
Hamiltonian constraint. For this aim, we calculate the following
terms respectively
\ba \left\{\ints \lambda(p-\phi\pi),\ints
\frac{2N}{\sqrt{h}}\left(\frac{p_{ab}p^{ab}-\frac12p^2}{\phi}\right)\right\}
&&=\ints\frac{N\lambda}{\sqrt{h}}\left(\frac{p_{ab}p^{ab}-\frac12p^2}{\phi}\right)\label{end1}
\ea

\ba \left\{\ints \lambda(p-\phi\pi),\ints
\frac{N\sqrt{h}}{2}\left(-\phi R\right)\right\}
&&=\ints\frac{N\lambda\sqrt{h}}{4}\left(-\phi R\right)-\ints
\lambda\sqrt{h}D_cD^c(\phi N) \ea

\ba \left\{\ints \lambda(p-\phi\pi),-\ints
\frac{3N\sqrt{h}}{4\phi}(D_a\phi)D^a\phi\right\}
&&=-\ints \frac{3N\lambda\sqrt{h}}{8\phi}(D_a\phi)D^a\phi+\ints
\frac{3\lambda\sqrt{h}}{2}D_a(ND^a\phi) \ea

\ba \left\{\ints \lambda(p-\phi\pi),-\ints
(D_aN)\sqrt{h}D^a\phi\right\} &&=\ints
\frac{\lambda}{2}(D_aN)\sqrt{h}D^a\phi+\ints\lambda\sqrt{h}\phi(D_aD^aN)
\ea

\ba \left\{\ints \lambda(p-\phi\pi),\ints N\sqrt{h}\xi(\phi)\right\}
&&=\ints\left(-\frac32\lambda N\sqrt{h}\xi(\phi)+\lambda\phi
N\sqrt{h}\xi'(\phi)\right) \label{end5}\ea Combining Eqs.
(\ref{end1})-(\ref{end5}), we have \ba
\{S(\lambda),H(M)\}=H(\frac{\lambda M}{2})+\ints
N\lambda\sqrt{h}(-2\xi(\phi)+\phi\xi'(\phi)). \ea

\end{document}